\documentclass[paper,11pt]{JHEP}

\usepackage[centertags]{amsmath}
\usepackage{amsfonts}
\usepackage{amssymb}
\usepackage{amsthm}
\usepackage{graphicx}

       \let\C=\Chi    

\newcommand{\C}{{\mathbb C}}
\newcommand{\be}{\begin{equation}}
\newcommand{\eeq}{\end{equation}}
\newcommand{\bea}{\begin{eqnarray}}
\newcommand{\eea}{\end{eqnarray}}
\newcommand{\ba}{\begin{array}}
\newcommand{\ea}{\end{array}}
\def\nn{\nonumber}
\newcommand{\ft}[2]{{\textstyle\frac{#1}{#2}}}

\newcommand{\ee}{\end{equation} }

\newcommand{\cP}{{\mathcal{P}}}

\newcommand{\one}{{\rm 1\kern -.9mm l}}

\newdimen\tableauside\tableauside=1.0ex
\newdimen\tableaurule\tableaurule=0.4pt
\newdimen\tableaustep

\def\phantomhrule#1{\hbox{\vbox to0pt{\hrule height\tableaurule
width#1\vss}}}
\def\phantomvrule#1{\vbox{\hbox to0pt{\vrule width\tableaurule
height#1\hss}}}
\def\sqr{\vbox{%
 \phantomhrule\tableaustep
\hbox{\phantomvrule\tableaustep\kern\tableaustep\phantomvrule\tableaustep}%
 \hbox{\vbox{\phantomhrule\tableauside}\kern-\tableaurule}}}
\def\squares#1{\hbox{\count0=#1\noindent\loop\sqr
 \advance\count0 by-1 \ifnum\count0>0\repeat}}
\def\tableau#1{\vcenter{\offinterlineskip
 \tableaustep=\tableauside\advance\tableaustep by-\tableaurule
 \kern\normallineskip\hbox
   {\kern\normallineskip\vbox
     {\gettableau#1 0 }%
    \kern\normallineskip\kern\tableaurule}%
 \kern\normallineskip\kern\tableaurule}}
\def\gettableau#1 {\ifnum#1=0\let\next=\null\else
 \squares{#1}\let\next=\gettableau\fi\next}
\newcommand{\Yfund}{\tableau{1}}

\title{Deformed Seiberg-Witten Curves for ADE Quivers}
\author{ Francesco Fucito,  Jose~F. Morales, Daniel Ricci Pacifici
\\
\vskip 0.2cm
I.N.F.N. - Sezione di Roma 2\\
and Universit\`a di Roma Tor Vergata, Dipartimento di Fisica\\
Via della Ricerca Scientifica, I-00133 Roma, Italy\\
\vskip 0.2cm

\vspace{0.25cm}
\email{Francesco.Fucito,Francisco.Morales,Daniel.Ricci.Pacifici@roma2.infn.it}
}
\abstract{We derive Seiberg-Witten like equations encoding the dynamics of ${\cal N}=2$ ADE quiver
gauge theories in presence of a  non-trivial $\Omega$-background along a two dimensional plane.  The $\epsilon$-deformed
prepotential and the chiral correlators of the gauge theory are extracted from difference equations that can be thought
   as a non-commutative (or quantum)  version of the Seiberg-Witten curves for the quiver.}

\preprint{ROM2F/2012/10}

\begin{document}

\tableofcontents
\section{Introduction}
It is well-known that the prepotential of a supersymmetric gauge theory with eight supercharges
can be computed exactly both at the perturbative and non-perturbative level \cite{Seiberg:1994rs,Seiberg:1994aj}. These results are elegantly
encoded in the geometry of Riemann surfaces.
While the original computations were carried out for theories with unitary gauge groups and fundamental or adjoint matter,
subsequent works extended those results to account for
orthogonal/symplectic groups, quiver (product of groups) gauge theories  and   symmetric or antisymmetric
matter \cite{Landsteiner:1997vd}\nocite{Landsteiner:1997ei,Ennes:1998ve,Ennes:1998gh,Ennes:1999nz,Ennes:1999rn}-\cite{Edelstein:1999dd} (see also \cite{Naculich:2004nm}
for a review and references). These extensions were possible once the original results were cast in the language of
string and M-theory \cite{Hanany:1996ie,Witten:1997sc}.

There is an alternative way, pioneered by Nekrasov and Okounkov \cite{Nekrasov:2003rj}, to recover these results
from a direct computation in the  gauge theory.  It is based on the evaluation of the partition function of the theory by means
of localization. Before being amenable to such treatment, the partition function $Z$, which is an integral over the instanton moduli space of
the gauge theory of interest, must be suitably deformed by
introducing two deformation parameters $\epsilon_\ell$ which break the Lorentz symmetries and regularize the spacetime volume
\cite{Nekrasov:2002qd}. Equivalently, one considers the gauge theory on a curved spacetime, the so called
 $\Omega$-background. For non-trivial $\epsilon_\ell$, the integral defining $Z(\epsilon_\ell,q)$ localizes around a set of isolated points allowing
for its explicit evaluation as a series in the instanton
winding number \cite{Nekrasov:2002qd}\nocite{Flume:2002az,Bruzzo:2002xf,Flume:2004rp,Nekrasov:2004vw,Marino:2004cn}--\cite{Fucito:2004gi} (see \cite{Moore:1998et}
for earlier applications of these ideas).
 The prepotential ${\cal F}$ of the gauge theory  is then identified with the free energy of the system in the limit where the
 $\epsilon$-background is turned off.
 Sending to zero the  parameters $\epsilon_\ell$, the Seiberg-Witten geometry encoding the non perturbative data
emerges from the equations describing the saddle point of the partition function \cite{Nekrasov:2003rj}.

  The study of the gauge theory dynamics beyond the limit $\epsilon_\ell \to 0$ is also of physical interest.
  Corrections in $\epsilon_\ell$ to the prepotential describe interactions of
gauge theory fields and gravity and  summarize an infinite tower of topological string amplitudes \cite{Antoniadis:2010iq}. Moreover,
the  ${\cal N}=2$ partition function
 $Z(\epsilon_\ell,q)$ at finite $\epsilon_\ell$ has been recently related to conformal blocks of 2D
CFT's \cite{Gaiotto:2009we,Alday:2009aq}. At last, the prepotential for  the case where one of the $\epsilon_\ell$ is sent to zero
has been identified with the Yangian of quantum integrable systems \cite{Nekrasov:2009rc}. These results triggered a rapid development
of the field along several directions
(see \cite{Mironov:2009uv}\nocite{Mironov:2009dv,Mironov:2009ib,Maruyoshi:2010iu,Dorey:2011pa,Mironov:2011jn,Chen:2011sj,Bonelli:2011na,Huang:2011qx,Manabe:2012bq,Mironov:2012uh}--\cite{Huang:2012kn} for recent studies of the
gauge theory dynamics in the $\Omega$-background).

  In \cite{Poghossian:2010pn,Fucito:2011pn} it was shown that a saddle point analysis can be adapted to the study of ${\cal N}=2$ $U(N)$ gauge
theories in presence of an $\Omega_\epsilon$-background  with $\epsilon_1=0$ and $\epsilon_2=\epsilon$.
 The exact prepotential and chiral correlators of the gauge theory in this limit were encoded in a  function $y(x)$ solving a  Seiberg-Witten
like equation. Using a different approach, similar results were also recently obtained in \cite{Hellerman:2011mv,Hellerman:2012zf}.

In this paper, we  apply a similar analysis to the case of ${\cal N}=2$ quiver gauge theories. We restrict ourselves
to asymptotically free  theories with unitary groups and fundamental matter.  The quiver gauge theories in this class are  given by
  taking as the quiver diagram  an oriented Dynkin diagram of an (Affine or not)  ADE Lie algebra. To each node we associate
   two integers $(N_a,n_a)$ characterizing the ranks of the gauge groups and the number of fundamentals.
  The  arrows in the quiver diagram label bifundamental matter.
The study of these quiver gauge theories and the derivation of the corresponding Seiberg-Witten curves governing the dynamics has been recently announced in \cite{NP}. The techniques in this paper
give an alternative derivation of these results and a generalization to the case of a non-trivial
  $\Omega_\epsilon$-background. Using matrix model techniques the case of quivers in the  $\Omega$-background has also been previously studied in \cite{Itoyama:2009sc}.

  The paper is organized as follows. In section \ref{squiver} we introduce the gauge theory models. In section
\ref{sinstanton} we describe  the instanton moduli spaces and the saddle point equations determining the leading
contribution to the partition function in the limit $\epsilon_1\to 0$. The saddle point equations are given by an infinite set of
conditions on the finite set (one for each gauge group) of functions $y_a(x)$  encoding the prepotential and chiral correlators of
the theory.  The set of saddle point equations will be summarized in section \ref{sdeformed} as a coupled system of polynomial
functional equations for $y_a(x)$ which reduce to the the Seiberg-Witten curves derived in \cite{NP} when the
$\epsilon$-background is turned off.  In section \ref{schiral} we show how chiral correlators of the gauge theory are computed out of
the $\epsilon$-deformed Seiberg-Witten differentials.
In section \ref{sdifferent}, we present alternative ways of writing the  $\epsilon$-deformed Seiberg-Witten equations
as a decoupled system of polynomial equations, as a quantum version of Seiberg-Witten curves and as a Thermodynamic Bethe ansatz like integral form. In appendix \ref{saffine} we collect some details of the Seiberg-Witten equations for the affine $A_1$-quiver gauge theory.

\section{Quiver gauge theories}
\label{squiver}

We consider asymptotically free quiver gauge theories with unitary gauge groups and (bi)fundamental matter. A quiver gauge theory
in this class is described in terms of a quiver diagram isomorphic to (an orientation of)
the Dynkin diagram of an (Affine or not) ADE algebra. To each node ``a", we associate a gauge group $U(N_a)$ and $n_a$ hypermultiplets in the fundamental representation
$\Yfund_a$ of $U(N_a)$. An arrow connecting nodes ``a" and ``b" describes a hypermultiplet in the bifundamental representation
  $(\Yfund_a,\overline{\Yfund}_b)$ with mass $m_{ab}$.  The hypermultiplet  content will be written as
\be
  {\cal H}_{\rm matter} =  \sum_{a,b} c_{ab}   (\Yfund_a,\overline{\Yfund}_b)  +\sum_a  n_{a}  \Yfund_a \label{matter}
\ee
 with $c_{ab}=0,1$ counting the number of arrows starting at node $a$ and ending on $b$. We stress the fact that, even if the
 quiver will be in general described by an orientation of the Dynkin diagram, i.e $c_{ab}\neq c_{ba}$,
  different orientations of the diagram are physically equivalent  since the states in the ${\cal N}=2$ hypermultiplets come always
  in CPT conjugated pairs. A flip in the orientation of an arrow can indeed, as we will see, be always reabsorbed in a redefinition of the
bifundamental masses.

In our study we will limit ourselves to the conformal case since the non-conformal set ups can be obtained from the former by sending some masses to
infinity. Conformal invariance translates into the condition
\be
\beta_a= -2 N_a+(c_{ab} +c_{ba}) N_b+n_a  = 0   \label{free}
\ee
which determines the number of fundamentals $n_a$ at each node in terms of the gauge group ranks $N_a$.
For $n_a=0$, the complete list  of conformal quiver gauge theories is
given by the  affine ADE quivers given in figure 1.

\begin{figure}[t!]
\includegraphics[width=8 cm]{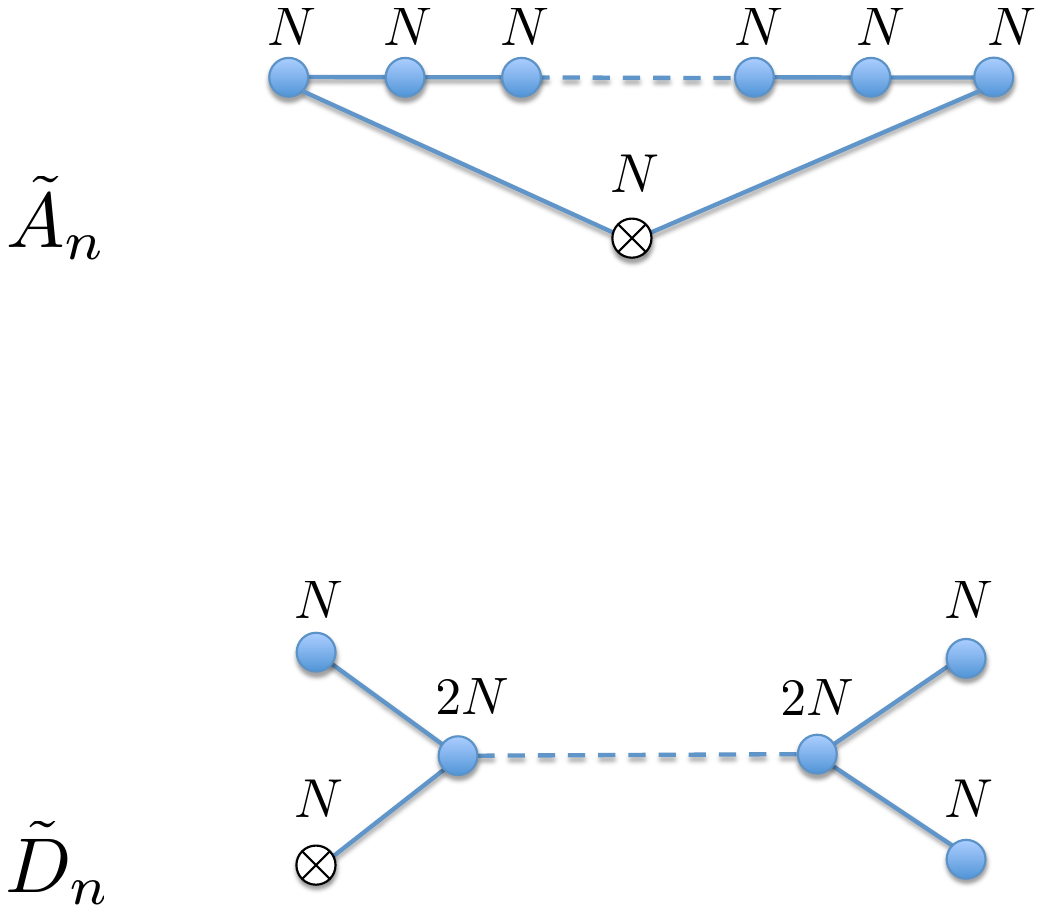}
\includegraphics[width=8 cm]{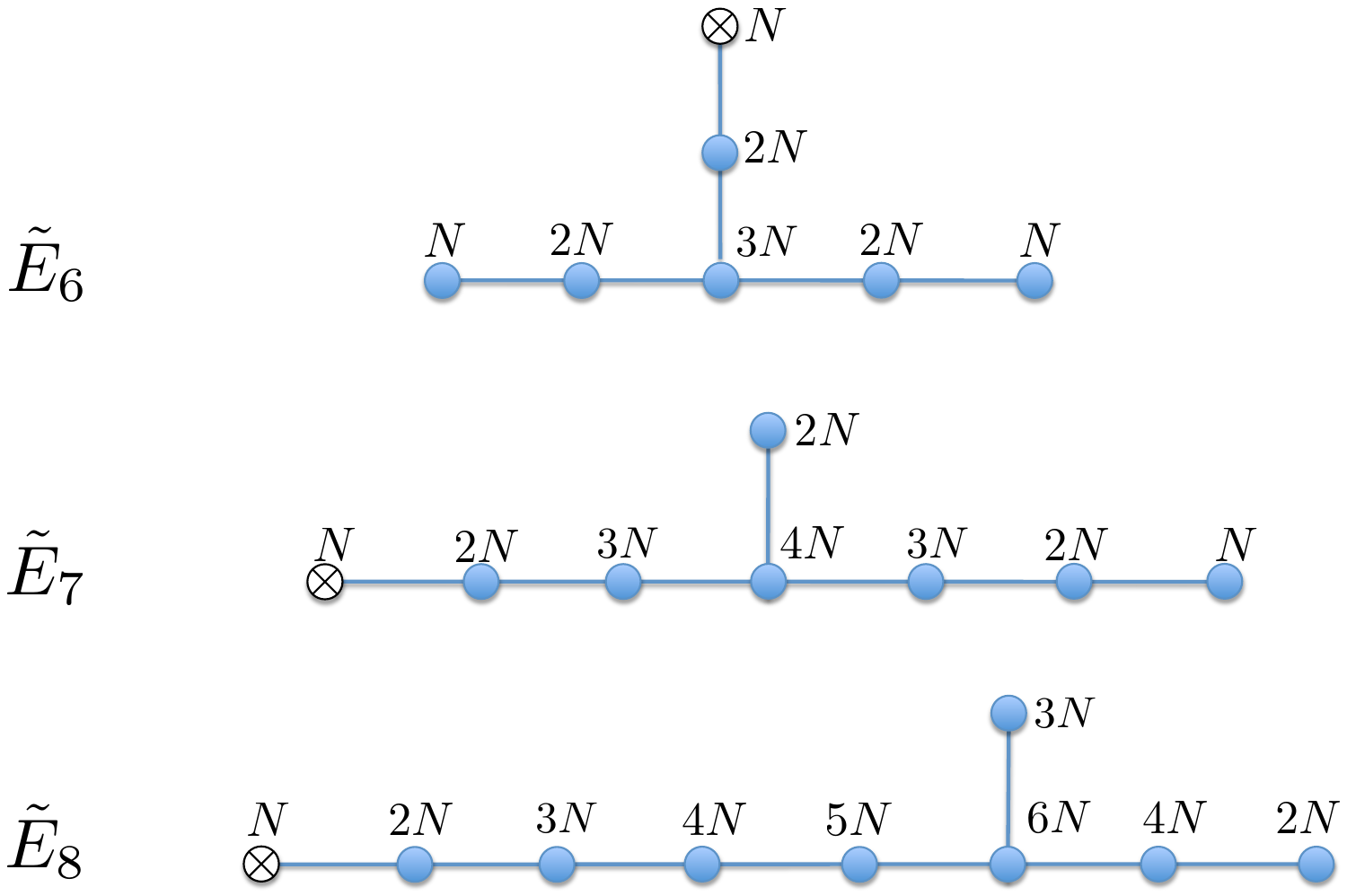}
\caption{Affine ADE Dynkin diagrams. Non-affine ADE diagrams are obtained by discarding the crossed node. The nodes are labelled by the rank
of the corresponding gauge group.}
\end{figure}

There are several realizations of the quiver gauge theories in string theory. The non-affine $A$-series is realized in type
IIA string theory by suspending D4 branes
between a sequence of NS5 branes  distributed along a line \cite{Hanany:1996ie,Witten:1997sc}. The nodes label the intervals
between the a$^{\rm th}$
and (a+1)$^{\rm th}$ NS5 brane. $N_a$ will be the numbers
of D4 branes stretched along this interval.  The open strings connecting the D4 branes inside the interval
  realize the $U(N_a)$ gauge degrees of freedom while the open strings stretched between the a$^{\rm th}$ and (a+1)$^{\rm th}$
  or (a-1)$^{\rm th}$ stacks lead to bifundamental matter. Finally fundamental matter is included by introducing  $n_a$ D6 branes.
The $A$-quiver diagram  corresponds to the choice $c_{ab}=\delta_{a,b-1}$.
Compactifying the  brane system   on a circle one finds the closed quiver diagram associated to the Dynkin diagram of
the Affine $\tilde A$ Lie algebra. At strong coupling, the brane system lifts to eleven dimensions  and can be described
in terms of M5 branes wrapping an $A_{r-1}$ orbifold singularity $\C^2/Z_r$. Similarly, the D and E series can be described in terms of an M5 brane wrapping the
corresponding ADE singularity $\C^2/\Gamma_{ADE}$, with $\Gamma_{ADE}$ a discrete subgroup of $SU(2)$.
 Alternatively, the  quiver gauge theory can be realized in terms of
fractional D3 branes at a $\C^2/\Gamma_{ADE}$ singularity. Fundamental matter in this setting can be included  by adding D7 branes.

  \section{Instanton moduli spaces}
\label{sinstanton}

In this section we describe the instanton moduli space of the ${\cal N}=2$ quiver gauge theories under discussion.
We refer to \cite{Bruzzo:2002xf,Fucito:2004gi,Billo:2012st} for a more detailed and a self-contained exposition.
We adopt the language of  fractional D3-branes (and flavor D7-branes)  to describe the quiver gauge theory. In this framework, instantons are
viewed as D(-1)-branes. Each node of the quiver corresponds to a type of fractional brane. We consider a general system of $k_a$, $N_a$
and $n_a$ D(-1), D3 and D7 fractional branes respectively.

Instanton moduli are in correspondence with the massless modes of open strings with at least one end on the D(-1)-branes.
The computation of observables in the gauge theory requires the evaluation of integrals over the moduli space spanned by these modes.
After a suitable deformation, these integrals localize around a finite set of critical
points \cite{Nekrasov:2002qd,Flume:2002az,Bruzzo:2002xf} allowing for their
explicit evaluation.   Complete localization requires the presence of an $\Omega$ background breaking the Lorentz symmetries. Moreover,
D-branes have to be distributed  along the transverse complex plane with no superposition in such
  a way that the full symmetry group $\prod_a U(k_a)\times U(N_a)\times U(n_a)$ is broken to its Cartan subgroup.
  We parametrize by $\chi_{I_a}$, $a_{u_a}$ and $m_{s_a}$ the Cartan elements of $U(k_a)$, $U(N_a)$ and $ U(n_a)$
  respectively. On the other hand $\epsilon_1$ and $\epsilon_2$ parametrize the Cartan of the Lorentz group.
 Geometrically, $\chi_{I_a}$, $a_{u_a}$ and $m_{s_a}$ specify the positions of D(-1), D3 and D7 branes respectively along the
 overall transverse plane.
The parameters  $a_{u_a}$, $m_{s_a},\epsilon_1$ and $\epsilon_2$ are part of the gauge theory data, while $\chi_{I_a}$ being an instanton modulus
should be integrated over.

 It is convenient to encode the symmetry data in the fundamental characters
  \bea
V_a=\sum_{I_a}^{k_a}  e^{i \chi_{I_a}} \qquad W_a=\sum_{u_a}^{N_a}   e^{i a_{u_a}} \qquad
W_{F,a}=\sum_{s_a=1}^{n_a} e^{i m_{s_a}  }
\eea
 and think of open strings connecting two branes as products of two of these functions.
For instance $V_a V_b^*$ represents an open string starting from a D(-1) brane of type ``a" and ending on a D(-1) brane of type ``b",
 $V_a W_b^*$ connects a D(-1) and a D3 brane, and so on.
 Finally we encode the Lorentz transformation properties of the fields by $T_\ell=e^{i \epsilon_\ell}$ .

Localization is based on the existence of a BRST charge ${\cal Q}$.  The instanton moduli organize into ${\cal Q}$-pair
$(\mathfrak M,\mathfrak N)$ related by
\be
{\cal Q}\, \mathfrak M=\mathfrak N      \qquad {\cal Q}^2\, \mathfrak M=\lambda\,  \mathfrak M
\ee
 with $\lambda$ the eigenvalue of the field $\mathfrak M$ with respect to the action of an element of the Cartan symmetry
 group parametrized by $\chi_{I_a}, a_{u_a}, m_{s_a}$ and $\epsilon_\ell$.
  The spectrum of eigenvalues $\lambda$ is summarized in the character
 \be
 {\bf T}={\rm tr}_{\mathfrak M} e^{i {\cal Q}^2}   \label{charT}
 \ee
  In table 1 we display the list of moduli and their contributions to   $ {\bf T}$.

 \begin{table}[h]
\begin{equation*}
\begin{array}{|c|c|c|}
\hline
\mathfrak M &  {\bf T} \\
\hline
\hline
 B_\ell,\lambda_c  & V_a\, V^*_a \, (T_1+T_2) -V_a\, V^*_a \, (1+T_1T_2)\\
   w,\bar{w}   & W_a \, V_a^* +V_a\, W_a^*\, T_1 T_2\\
   \hline
    B_{\dot \ell},\lambda_{m'}  &c_{ab} V_a\, V^*_b \, (1+T_1T_2)  -c_{ab} V_a\, V^*_b \, (T_1+T_2) \\
 \mu,\bar{\mu}   & -c_{ab} W_a \, V_b^* -c_{ab} V_a\, W_b^*\, T_1 T_2\\
\hline
\mu' &  -V_a W^*_a   \\
\hline
\end{array}
\end{equation*}
\label{modulitable}
\caption{Instanton moduli for ${\mathcal N}=2$ quiver gauge theories.
}
\end{table}

 The three main rows display the contributions coming from the moduli associated to gauge, bifundamental and fundamental matter degrees of freedom
 respectively.  The first column displays the highest weight states of the ${\cal Q}$-pairs and
  the second column their  contributions to the character  of the given field.
  In particular the fields $B_{\ell=1,2},B_{\dot\ell=3,4}$,  parametrize the positions of the instanton along the spacetime and the
 transverse space respectively.  $w,\bar{w}$ come from open strings stretching between D(-1) and D3 branes.
 Fermions $\lambda_c,\lambda_{m'},\mu,\bar \mu,\mu'$ contribute with a minus sign and  account for the implementation of the ADHM
 constraints which reduce the number of the degrees of freedom.

 \subsection{Instanton partition function}

    The instanton partition function is defined by
 \be
Z_{\rm inst}=\int d{\cal M} e^{-S_{D(-1)D3}}     \label{zinst}
\ee
  After the equivariant deformation this integral localizes around a set of isolated fixed points of ${\cal Q}^2$. These points are
  in one-to-one correspondence with
the arrays $Y=\{ Y_{u_a}  \}$ $u_a=1,..N_a$ of Young tableaux centered at the D3-brane positions $a_{u_a}$.
    Each box in the array of tableaux represents an instanton with position
 \be
\chi^Y_{I_a}=\chi^Y_{u_a,ij} =a_{u_a}+(i-1) \epsilon_1+ (j-1) \epsilon_2   \label{chiy1}
\ee
with $i,j$ running over the rows and columns of the tableau $Y_{u_a}$.
        The partition function reduces to
\be
Z_{\rm inst}= \sum_Y Z_Y= \sum_Y { q^{|Y|}\over {\rm Sdet}_Y {\cal Q}^2 }   \label{sdet}
  \ee
with  ${\rm Sdet}_Y {\cal Q}^2$  the  superdeterminant of ${\cal Q}^2$ evaluated at (\ref{chiy1}) for a given $Y$.
The eigenvalues entering into (\ref{sdet}) can be extracted from the  character $ {\bf T}$ defined in (\ref{charT}) and listed in table 1.
 Collecting all the contributions one finds
 \be
{\bf T}_Y  =\sum_{ab} t_{ab}   {\bf T}_{\rm ab}-T_F\label{ty}
\ee
with $t_{ab}= \delta_{ab}-c_{ab} e^{i m_{ab}}$,
and
 \bea
 {\bf T}_{\bf ab,Y}&=& -V_a\, V_b^* \, (1-T_1)(1-T_2)+W_a V^*_b  +V_a \, W^*_b \, T_1\, T_2      \nn\\
 T_{F,Y}&=& \sum_a V_a W^*_{F,a}
   \label{chiy}
\eea
Here the $V_a$'s are evaluated at the instanton positions (\ref{chiy1}) associated to the critical point described by $Y$.
The extra phases $m_{ab}$ parametrize the masses of bifundamentals.

Collecting the eigenvalues from (\ref{ty},\ref{chiy}) and taking $q^{|Y|}= \prod_a [(-)^{N_a-c_{ba} N_b}q_a]^{k_a}$  one
finds\footnote{One can easily check that
the number of zero eigenvalues in the numerator and denominator of this formula matches, leading to a finite result.}
\bea
Z_Y&=&  \prod_{a,I_a} \left[   -{  Q_a(\chi^Y_{I_a} ) \over
  P_{a}(\chi^Y_{I_a})  P_{a}(\chi_{I_a}^Y+\epsilon_1+\epsilon_2)  }
 \prod_{J_a} \Delta(\chi_{I_a}^Y-\chi^Y_{J_a})  \right] \nn\\ &&\times
 \prod_{a,b,I_a}   \left[  { P_{b}(\chi_{I_a}^Y-m_{ba} )^{c_{ba}}     P_{b}(\chi^Y_{I_a}+\epsilon_1+\epsilon_2+m_{ab} )^{c_{ab}}
   \over     \prod_{J_b} \Delta(\chi_{I_a}^Y-\chi^Y_{J_b}+m_{ab})^{c_{ab}}
   }   \right] \label{zz}
 \eea
 with
\bea
    \Delta(x)&=& {x (x+\epsilon_1+\epsilon_2)\over (x+\epsilon_1) (x+\epsilon_2)}
     \qquad    P_{b}(x)= \prod_{u_b=1}^{N_b} (x-a_{u_b} ) \nn\\
  Q_a(x) &=&q_a  \prod_{s_a=1}^{n_a} (x-m_{s_a})    \qquad
  \label{QP}
\eea
    We notice that a flip in the orientation of an arrow in the quiver diagram sends $c_{ab} \leftrightarrow c_{ba}$
and it can be reabsorbed in the redefinition $m_{ba} \leftrightarrow -m_{ab}-\epsilon_1-\epsilon_2$.
The prepotential of the gauge theory is given by
\be
{\cal F}_{\rm inst}(\epsilon_\ell,a,q)=-\epsilon_1\epsilon_2\log Z_{\rm inst}(\epsilon_\ell,a,q)
\label{prepotential}\ee
The chiral correlators are computed by the localization formula
\be
\langle {\rm tr}_{N_a} e^{iz\,  \Phi} \rangle =\sum_{u_a=1}^{N_a} e^{i z a_{u_a}}  -
(1-T_1^z)(1-T_2^z){1 \over Z_{\rm inst}} \sum_Y Z_Y\,  \sum_{I_a} \,e^{iz\,\chi_{I_a}}
\label{trfij}
\ee

\subsection{Saddle point equations}

The instanton partition function is
\bea
Z_{\rm inst}=\sum_{ Y} e^{\ln Z_Y}
\eea
with the sum running over the arrays of $\sum_a N_a$ Young tableaux.
From (\ref{zz}) one finds
\be
\ln Z_Y=\sum_{a,b}  \sum_{ I_a, J_b }  F_{ab} (\chi^Y_{I_a}-\chi^Y_{J_b})+\sum_{a}
 \sum_{I_a }  M_{a} (\chi^Y_{I_a})
 \ee
with
\bea
F_{ab} (x) &=& \delta_{ab}   \ln \Delta(x)  -c_{ab} \ln \Delta(x+m_{ab} )    \\
M_{a} (x)&=& \ln \left(  - { Q_a(x)\over  P_{a}(x)P_{a}(x+\epsilon_1+\epsilon_2)   } \right) \nn\\
&&+\sum_b \left[c_{ab} \ln P_{b}(x+m_{ab}+\epsilon_1+\epsilon_2)
+  c_{ba} \ln  P_{b}(x-m_{ba} )  \right]\nn
\eea

We are interested in the limit $\epsilon_1\to 0$, keeping $\epsilon_2=\epsilon$ finite.   We will follow the strategy in \cite{Poghossian:2010pn,Fucito:2011pn} where a similar analysis has been performed for $U(N)$ gauge theories with fundamental and adjoint matter. We refer the reader to these references for further details.
In the limit $\epsilon_1\to 0$, the instantons of type ``a" form a continuous  distribution along the intervals
\be
{\cal I}_a =\cup_{u_a} {\cal I}_{u_a} =\cup_{u_a i} \, [x^0_{u_a i},x_{u_a i} ] \label{ia}
\ee
 with
 \be
 x^0_{u_a,i}=a_{u_a}+ (i-1) \epsilon
 \ee
 $x_{u_a,i}$ parametrizes the height of the $i^{\rm th}$ columns in the tableau.  The details
 of the Young tableau can be encoded in the instanton density
 \be
 \rho_a(x) =\epsilon_1 \sum_{I_a} \delta(x-\chi_{I_a})=\left\{
  \begin{array}{cc}
  1 & x \in   {\cal I}_a \\
  0 & x \notin   {\cal I}_a
\end{array}\right.
\label{rho}
 \ee
 We write
 \be
\ln Z_Y={1\over \epsilon_1} {\cal H}_{\rm inst}(\rho)
\ee
with
\be
{\cal H}_{\rm inst}(\rho) =\ft12 \sum_{a,b}  \int dx dy \rho_a(x) \rho_b(y)   G_{ab} (x-y)+
 \sum_a \int dx   \rho_a(x)  M_{a} (x)  \label{hh}
 \ee
 and
 \be
 G_{ab}(x)= \lim_{\epsilon_1\to 0} {1\over \epsilon_1} \left[ F_{ab}(x)+F_{ba}(-x)\right]
  \ee
   The partition function becomes
\be
Z_{\rm inst}=
 \int D\rho\, {\rm e}^{ {1\over \epsilon_1} {\cal H}_{\rm inst}(\rho)}   \label{zr2}
 \ee
 In the limit $\epsilon_1\to 0$ the main contribution to the integral comes from the  instanton distribution $\rho(x)$
 extremizing the Hamiltonian
 \be
 {\delta\rho_a \over \delta x_{u_a i}}\, {\delta {\cal H}\over \delta \rho_a}
 = \sum_{b}  \int   dy   \rho_b(y)   G_{ab} (x_{u_a i}-y)+
  M_{a} (x_{u_a i})  =0   \label{saddle}
 \ee
 To perform the integral in (\ref{saddle}) it is convenient to write $G_{ab}(x)$ in the form
 \be
 G_{ab}(x)={d\over dx} \left[  \delta_{ab}\ln\left( {x+\epsilon\over x-\epsilon }\right)+
   c_{ab}\ln\left( {x+m_{ab} \over x+m_{ab}+\epsilon }\right)+
   c_{ba}\ln\left( {x-m_{ba}-\epsilon \over x-m_{ba} }\right)\right]   \label{gdlog}
 \ee
and introduce the function
\be
{\cal Y}_a(x)= \prod_{v_a=1}^{N_a} \prod_{i=1}^\infty \left({ x- x_{v_a i}  \over
  x- x^0_{v_a i}  } \right)
\label{yy}
\ee
 encoding the information about the $x_{u_a i}$.
Each term in the saddle point equation (\ref{saddle}) can be written in terms of this function using the identity
 \be
 \int  dy\,\rho_a(y)\, {d\over dx} \ln (x-y)=- \ln {\cal Y}_a(x)  \label{yy}
 \ee
The saddle point equations can be thought of as an extremization over the $\cal Y$-function.
 Alternatively, one can use the $x_{ui,a}^0$-independent
combinations\footnote{$y(x)$ is related to the variable $w(x)$ in \cite{Fucito:2011pn} as $y(x)=1/w(x)$.}
\be
  y_a(x)={   {\cal Y}_a(x) P_{a}(x) \over {\cal Y}_a(x-\epsilon)}=\prod_{v_a=1}^{N_a} \prod_{i=1}^\infty \left({ x- x_{v_a i}  \over
  x- x_{v_a i} -\epsilon } \right)
 \label{yw}
 \ee
 We notice that at large $x$ ${\cal Y}_a\approx 1$ and
 \be
 y_a \approx  x^{N_a}   \label{yxlarge}
 \ee
In terms of this function the saddle point equation (\ref{saddle}) can be written in the compact form
\be
 \framebox[1.15\width ][c]{$
  1+Q_a(x ) {\prod_b y_b(x  -m_{ba} )^{c_{ba}} \,y_b(x +m_{ab}  +\epsilon)^{c_{ab}}
\over y_a(x  )\,y_a(x  +\epsilon)}=0     \quad {\rm for} \quad x\in \{  x_{u_a i} \}
  $}  \label{eqsad}
\ee

\section{Deformed Seiberg-Witten equations }
\label{sdeformed}

In this section we show that the saddle point equations (\ref{eqsad}) can be equivalently written as
  Seiberg-Witten like equations for the functions $y_a(x)$.
   We  will follow again the strategy of  \cite{Poghossian:2010pn,Fucito:2011pn} adapting the analysis to the quiver theory.
   The main idea is to exploit the saddle point equations to build a set of rational functions, $\chi_a$, of the $y$'s
with no poles in the complex $x$-plane, i.e. a set of polynomials.
In this way, $\chi_a$ is completely determined in terms of a finite number of coefficients.

\subsection{The $\epsilon=0$ case}

We will start by considering the $\epsilon=0$ case with massless bifundamental hypermultiplets, i.e. $m_{ab}=0$, making
contact with the results  in \cite{NP}.
  In this limit, the Young tableaux profile becomes a smooth curve and the details
of the instanton
saddle point configuration are encoded in a set of continuous functions $y_a(x)$ with cuts in the $x$-plane.
The Seiberg-Witten equations were deduced \cite{NP} from a careful analysis of the discontinuities of these functions across the cuts.
We start by illustrating how these results can be recovered from the $\epsilon$-deformed saddle point equations (\ref{eqsad})
after turning off the $\epsilon$-background. The discussion will serve as a warm up for the   $\epsilon\neq 0$
analysis in the next subsection.

 First we notice that in the limit  $\epsilon \to 0$, the points $x_{u_a i}$ form a continuous distribution around the $e_{u_a}$'s
  filling the intervals ${\cal I}_{a}$ defined in (\ref{ia}).  It is convenient to introduce the following function
  \be
 f_a(x)=y_a(x) +Q_a(x ) {\prod_b y_b(x )^{c_{ab}+c_{ba} }\over y_a(x)}
   \label{eqsad0}
\ee
Using (\ref{eqsad}) one can  see that these functions have no poles around $  {\cal I}_{a} $.
Indeed, the two terms in the r.h.s. of (\ref{eqsad0}) have simple poles in $x_{u_a i}$,
but they cancel against each other according to  (\ref{eqsad}).
On the other hand the functions $f_a(x)$ have poles around  ${\cal I}_{b}$
with $b$ linked to
$a$ in the quiver diagram. Again these poles  can be canceled  by the replacement
\be
 y_b(x)   \to   f_b(x)
  \label{iweyl}
\ee
 in the second term in (\ref{eqsad0}). This replacement will however generate new poles at higher orders in the $Q_a$'s
  that can be canceled again by the substitution (\ref{iweyl}).
According to the case at study, the iteration process will close in a finite or infinite number of steps. Interestingly,
one can see that the  terms generated in this way match the weights of the
basic representations  ${\bf R}_a$ of the Lie algebra. More in detail, a term of the type $y_1^{p_1}y_2^{p_2}\ldots$
can be put in correspondence with a weight $(p_1,p_2,\ldots )$ of the representation  ${\bf R}_a$.
Therefore $\chi_a$ can be thought of as a $Q$-deformed
version of its character.
The function $\chi_a(y,Q)$ obtained in this way is in general a polynomial in the $Q_a$'s with coefficients given by ratios of $y_a$'s.
Moreover,  $\chi_a(y,Q)$ has no poles   in the complex plane and therefore is a polynomial $\cP_a(x)$.
Using (\ref{yxlarge}) one can easily see that all the terms in $\chi_a$ grow like $x^{N_a}$ at large $x$,  and
therefore $\cP_a$ is a polynomial of order $N_a$ in $x$.
 One can then write
  \be
 \framebox[1.15\width ][c]{$  \chi_a( y,Q)= \mathcal{P}_a(x)   $}    \label{defsw}
\ee
The system of equations (\ref{defsw}) summarizes the content of the infinite number of algebraic saddle point equations (\ref{eqsad}) in the limit of $\epsilon\to 0$ and generalizes the Seiberg-Witten equations to the case of a quiver gauge theory.
The equations (\ref{defsw}) can be solved for $y_a(x)$  in terms of the polynomials  $\cP_a$. The prepotential and chiral correlators of the quiver gauge theory will be computed in the next section in terms of the periods of the
 $\epsilon$-deformed Seiberg-Witten differentials built out of the $y_a$'s.

The result obtained by the iterative procedure (\ref{iweyl})  can be alternatively obtained iterating the maps
 \be
s_a:  y_b(x) \to \left \{  \begin{array}{ll}
a=b  &   Q_a(x )\, y_a(x) \,  \prod_b y_b(x )^{-C_{ab} } \\
 a\neq b  &  y_b(x)    \\
\end{array}
\right.
 \label{iweyl0}
 \ee
with
\be
C_{ab}=2 \delta_{ab}-c_{ab}-c_{ba} \label{cartan}
\ee
the Cartan matrix of the ADE Lie algebra.  The map $s_a$  is a sort of Q-deformed Weyl reflection.
Indeed, acting with $s_a$ on $y_a^{-1}$ in the denominator brings you back to the previous step in the recursion, i.e. $s_a^2=1$.  The group $W$ generated by the $s_a$'s is isomorphic to the Weyl group of the corresponding algebra
and was named the iWeyl group in \cite{NP}.  We refer to the orbit starting from $y_a$ as the iWeyl orbit.

It is important to notice that for some representations of the D and E series higher powers of $y_a^n$ arise in the iWeyl
orbit and the replacement $y_a^n \to f_a^n$ cannot be realized as a result of an iWeyl reflection. In these cases,
the extra terms in $f_a^n$ generate new iWeyl orbits completing the weights of the associated representation.
For example, for $D_4$ the  ${\bf 28}$ leads to a length $24$ orbit and four  singlets,  in the $D_5$ for the ${\bf 120}$  one finds a length $80$
and four length ${\bf 10}$ orbits, for the ${\bf 45}$ we get a $40$ plus five singlets and so on and so forth.
On the other hand for the A series all
characters $\chi_a$ are made out of a single orbit.

 We write
  \be
  \chi_a(x)=\sum_{w\in {\rm Orb}_a} w(y_a)
  \ee
   with ${\rm Orb}_a$ the collection of iWeyl orbits spanning all weights of the representation ${\bf R}_a$.

    For instance, for $A_3$ one finds   the recursion trees
      \bea
      \label{trees}
\begin{array}{cccccccccccc}
& &~& y_1  &\buildrel s_1 \over{\rightarrow}  &  {Q_1  y_2\over y_1 }  & \buildrel s_2 \over{\rightarrow} &   {Q_1 Q_2  y_3\over y_2 }
& \buildrel s_3 \over{\rightarrow}  & {Q_1 Q_2 Q_3  \over y_3 } & \\ \\
 &&& &&&  \buildrel s_1\over{} \nearrow & {Q_1 Q_2  y_3\over y_1} & \searrow  \buildrel s_3 \over{} &   \\
A_3:& \bigcirc\to\bigcirc\to\bigcirc& &     y_2 &\buildrel s_2 \over{\rightarrow} &   {Q_2  y_1 y_3\over y_2 }  &   & & &
{Q_1 Q_2 Q_3  y_2\over y_1 y_3} & \buildrel s_2 \over{\rightarrow}&  {Q_1 Q_2^2  Q_3 \over y_2}\\
&&&  &&&  \buildrel s_3 \over{} \searrow &    { Q_2 Q_3  y_1\over y_3}  &     \buildrel s_1\over{}   \nearrow &   \\\\
 &  & & y_3  &\buildrel s_3 \over{\rightarrow}  &  {Q_3  y_2\over y_3 }  &\buildrel s_2 \over{\rightarrow} &  {Q_2 Q_3  y_1\over y_2 }
&\buildrel s_1 \over{\rightarrow} & {Q_1 Q_2 Q_3  \over y_1 } & \\ \\
\end{array}
\eea
 leading to the characters
\bea
\chi_1 &=&  y_1  + {Q_1  y_2\over y_1 } + {Q_1 Q_2   y_3\over y_2 } + {Q_1 Q_2 Q_3  \over y_3 } \nn\\
\chi_2  &=&   y_2 +  {Q_2   y_1 y_3\over y_2 }  +  {Q_2 Q_3   y_1\over y_3 } +{Q_1 Q_2   y_3\over y_1}  +
 {Q_1 Q_2 Q_3   y_2\over y_1 y_3} +  {Q_1 Q_2^2  Q_3 \over y_2}\nn\\
 \chi_3 &=&  y_3  +  {Q_3  y_2\over y_3 }  +{Q_2 Q_3  y_1\over y_2 } + {Q_1 Q_2 Q_3  \over y_1 } \label{a3}
\eea
 The Seiberg-Witten curves follow from equating $\chi_a$ in (\ref{a3}) to $\cP_a$. The same results were recently obtained in
\cite{NP} by a careful study of the discontinuities of functions $y_a$ crossing the cuts ${\cal I}_a$ in the complex plane.

%

\subsection{Turning on the $\epsilon$-background }

 The analysis in the previous section can be easily adapted to the case of a non-trivial $\Omega$ background $\epsilon$
 and massive bifundamental matter.  Using the saddle point equations (\ref{eqsad})
we can build the function
  \be
 f_a(x)=  y_a(x) +   Q_a(x-\epsilon) {\prod_b y_b(x-m_{ba}-\epsilon  )^{c_{ba}} \,y_b(x +m_{ab} )^{c_{ab}}
\over  y_a(x-\epsilon)}   \label{fa}
    \ee
   with no poles around ${\cal I}_a$. Indeed, both terms in (\ref{fa}) have poles at $\{ x_{u_a i}+\epsilon \}$
   but they cancel against each  other as it follows from (\ref{eqsad}). As before, the function $f_a(x)$ has
   poles around ${\cal I}_b$  that can be   canceled by the replacement
       \be
    y_b(z) \to f_b(z)   \label{ytob}
    \ee
     in the numerator of the second term in (\ref{fa}). The process is iterated order by order in the $Q_i$'s and leads,
     as before, to a continuous function $\chi_a(z)$ with no poles in the complex plane, i.e. a polynomial.
   Alternatively, one can implement the sequence of replacements (\ref{ytob}) as a recursion  generated at
   each level by the
  map
      \be
     y_a(x)  \to  Q_a(x-\epsilon) {\prod_b y_b(x-m_{ba}-\epsilon  )^{c_{ba}} \,y_b(x +m_{ab} )^{c_{ab}}
 \over  y_a(x-\epsilon)}  \label{iweyl2}
    \ee
    which is a sort of ``quantum version" of   the iWeyl reflection (\ref{iweyl0}). The characters $\chi_a(x)$ are built from $y_a(x)$
    following the same steps as before (see (\ref{trees}) for example for $A_3$), with  the arguments of the $Q_a$'s and
    $y_a$'s shifted at each step according to (\ref{iweyl2}).
     Explicitly, for an $ A_2$-quiver with  $c_{12}=1$, $m_{12}=m$, the characters entering in the Seiberg-Witten
     equations are given by
      \bea
A_2: ~~ \chi_1 &=& y_1(x) +  Q_1(x-\epsilon) {  y_2(x +m)    \over  y_1(x-\epsilon)}+ { Q_1(x-\epsilon) Q_2(x+m-\epsilon)     \over
   y_2(x+m-\epsilon)} \nn\\
  \chi_2 &=& y_2(x) +   Q_2(x-\epsilon) {  y_1(x-m-\epsilon )    \over  y_2(x-\epsilon)}+ {Q_1(x-m-2\epsilon) Q_2(x-\epsilon)   \over
 y_1(x-m-2\epsilon)  }
 \eea
    The first two terms in $\chi_1$ have poles in $x_{u_1 i}$ that cancel against each other according to (\ref{eqsad}).
  A similar cancellation is achieved between the poles at $x_{u_2 i}$ of the  last two terms in $\chi_1$. The analysis for
  $\chi_2$ is identical.

   The deformed Seiberg-Witten equations  can be written as before as
  \be
 \framebox[1.15\width ][c]{$  \chi_a(y,Q)= \mathcal{P}_a(x)   $}   \label{defsw2}
\ee
  with the $\epsilon$-deformed characters $\chi_a(y,Q)$ given in terms of the recursion trees generated by (\ref{iweyl2}) starting
  from $y_a(x)$ and $\cP_a$ some polynomials of order $N_a$.

  \section{Chiral correlators}
  \label{schiral}

Let $y_a(x)$ be a solution of the deformed Seiberg-Witten equations (\ref{defsw2}).
The chiral correlators
of the $SU(N_a)$ gauge group can be computed according to \cite{Fucito:2011pn}
\be
\langle {\rm tr}_{N_a} \Phi^J \rangle
 =  \int_{\gamma_a} {dx\over 2\pi i}  \,x^J\,   \partial_x \ln
y_a(x)
\label{thphij0}
\ee
with $\gamma_a$ a cycle around    ${\cal I}_a$ defined in (\ref{ia}). At the instanton level $k$, this integral receives
 contributions from the poles of the form  $x=e_{u_a}+i \epsilon$ with $i=0,\ldots k$ and $e_{u_a}$ the zeroes of $\cP_a(x)$.
 More precisely, we write
\be
     \mathcal{P}_a(x) = g_a \prod_{u_a=1}^{N_a} (x-e_{u_a})
\ee
  with the $e_{u_a}$'s parametrizing the quantum moduli space of the gauge theory.
The coefficient  $g_a$ is  fixed by
matching the leading behavior at large $x$ of the two sides of the equation
  \be
  g_a=x^{-N_a} \lim_{x\to \infty}  \chi_a (x)=\chi_a (Q_a,y_a)\Big|_{Q_a,y_a\to 1}   \label{gg}
     \ee
    The expression for $g_a$ in the non-conformal case can be found from (\ref{gg}) by sending to zero all the $q$'s associated to a
non-conformal node.

The quantum parameters $e_{u_a}$ can be determined in terms of the classical vevs $a_{u_a}$
by inverting the relations
\be
a_{u_a}
 =  \int_{\gamma_{u_a}} {dx\over 2\pi i}  \,x\,   \partial_x \ln
y_a(x)   \label{aint}
\ee
with $\gamma_{u_a}$ a cycle round ${\cal I}_{u_a}$.
The non-perturbative prepotential  ${\mathcal F}$ of the gauge theory can be extracted from
the expression for   $\langle {\rm tr}_{N_a} \Phi^2 \rangle $ using the Matone relation \cite{Matone:1995rx,Matone:2002wh}
\be
\langle {\rm tr}_{N_a} \Phi^2 \rangle = \sum_{u_a=1}^{N_a}a_{u_a}^2+2\,q_a\, \frac{\partial {\mathcal F}}{\partial q_a}
\label{matone}
\ee

In the next subsection we will match the results obtained from (\ref{thphij0}) with those coming from the direct evaluation
of the instanton partition function (\ref{trfij}) at $k=1$ instanton level.
We have also performed higher instanton checks of the deformed Seiberg-Witten curves for various quivers of the ADE series.

  \subsection{ $k=1$}

  In this section we display the  $k=1$ instanton contribution for a general quiver.
  The deformed Seiberg-Witten equations  up to this order can be written as
 {\small
 \be
   \cP_a(x) = y_a(x) + Q_a(x-\epsilon){\prod_b y_b(x-\epsilon-m_{ba}  )^{c_{ba}} \,y_b(x+m_{ab} )^{c_{ab}}   \over  y_a(x-\epsilon)}+\ldots
   \label{affcase}
   \ee
}
 The solution of the system of equations (\ref{affcase})  at first order in the $q_a$ reads
  \be
  y_a(x) = \cP_a(x)- Q_a(x-\epsilon)\, {\prod_b \cP_b(x-\epsilon-m_{ba}  )^{c_{ba}} \,\cP_b(x+m_{ab} )^{c_{ab}}   \over  \cP_a(x-\epsilon)}+\ldots    \label{yya}
  \ee
  Plugging this into (\ref{aint})  one finds
    \bea
a_{u_b} &=&  \sum_{i=0}^1 {\rm Res}_{z=e_{u_b}+i\,\epsilon}\,  x\,\partial_x  \ln
y_b(x)+\ldots \nn\\
&=& e_{u_b} +{1  \over {P}'_b(e_{u_b})} \,\left[
 Q_b(e_{u_b}  ) {\prod_d  { P}_d(e_{u_b} -m_{db})^{c_{db}} { P}_d(e_{u_b }+ \epsilon+m_{bd})
^{c_{bd}}  \over  {P}_b(e_{u_b} +\epsilon)} \right. \nn\\
 &&\left.
+    Q_b(e_{u_b} - \epsilon) {\prod_d  { P}_d(e_{u_b}-  \epsilon-m_{db})^{c_{db}} { P}_d(e_{u_b,b}+m_{bd})
^{c_{bd}}  \over  {P}_b(e_{u_b}- \epsilon)}     \right]+    \ldots
\eea
which can be simply inverted and used to evaluate
\bea
&& \langle {\rm tr} \Phi^J \rangle_{b}= \sum_{i=0}^1\sum_{u_b=1}^{N_b} {\rm Res}_{z=e_{u_b}+i\,\epsilon}\,  x^J\,\partial_x  \ln
y_b(x)+\ldots =\sum_{u_b=1}^{N_b}a_{u_b} ^J \\
&& + J   \,\,\sum_{u_b=1}^{N_b} \left[
 (a_{u_b}+\epsilon)^{J-1} -  a_{u_b}^{J-1} \right]
   Q_b(a_{u_b}  ) {\prod_d  { P}_d(a_{u_b} -m_{db})^{c_{db}} { P}_d(a_{u_b }+ \epsilon+m_{bd})
^{c_{bd}}  \over   {P}_b'(a_{u_b}) { P}_b(a_{u_b} +\epsilon)} +    \ldots \nn
 \eea
  In particular, using the  Matone relation (\ref{matone}) one finds the first instanton order to the quiver gauge theory
  prepotential
  \be
  {\cal F}_{\rm inst}=\sum_{b}\sum_{u_b=1}^{N_b}   \epsilon\,    \,\,    Q_b(a_{u_b}  ) {\prod_d  { P}_d(a_{u_b} -m_{db})^{c_{db}} {P}_d(a_{u_b }+ \epsilon+m_{bd})
^{c_{bd}}  \over   { P}'(a_{u_b}) {P}_b(a_{u_b} +\epsilon)} +    \ldots   \label{ff}
  \ee
  This formula matches the result for the one-instanton contribution to ${\cal F}_1$ coming from (\ref{prepotential})
  \bea
  {\cal F}_1=-\epsilon_1 \epsilon_2 Z_1= -\epsilon_1 \epsilon_2 \sum_b \sum_{u_b}  Z_{\Yfund_{u_b}}
    \eea
 with $Z_{\Yfund_{u_b}} $ given by (\ref{zz}) with $\chi_{I_b}=a_{u_b}$, $\epsilon_1=0$ and $\epsilon_2=\epsilon$.

\section{ Alternative formulations of Seiberg-Witten equations}
\label{sdifferent}

\subsection{ Seiberg-Witten equations in the non-commutative space}

The Seiberg-Witten equations (\ref{defsw}) for a quiver gauge theory at $\epsilon=0$ can be alternatively written in terms of the set of polynomial equations \cite{NP}
\be
{\rm det}_{\bf R_a} (\one-y_a^{-1}{\bf g} )\Big|_{\chi_b\to \cP_b}=0
\label{detga}
\ee
with ${\bf g}={\rm diag} \{ \chi_{a,i} \} $ a diagonal matrix with entries the components of $\chi_a(x)$. This implies in particular that $\chi_a(x)={\rm tr}\,{\bf g}$. We remark that the system of equations  (\ref{detga})
is only apparently decoupled since the polynomials appearing in  (\ref{detga}) depend on all the quantum variables ${e_{u_a}}$. The evaluation of the chiral correlators of the $U(N_a)$ gauge theory
requires the knowledge of all the periods   $a_{u_a}$  (and therefore of all the $y_a(x)$) in order to be able to find
  the relation between the quantum $e_{u_a}$ and the classical variables $a_{u_a}$ computed from (\ref{aint}).

In this section we show how these results can be extended to the $\epsilon \neq 0$ case for  the  simplest case, the fundamental representation  ${\bf R_1}$ of the $A_{r-1}$ quiver \footnote{We thank R. Poghossian for discussions on this point.}.
 We will then show how the results can be reinterpreted as a non-commutative (or quantum) version of the Seiberg-Witten curves for the quiver.

  By construction
\be
0={\rm det}_{\bf R_1} (\one-y_1^{-1}{\bf g} )=   \sum_{j=0}^r   \sum_{i_1<\ldots  <i_j}  \prod_{p=1}^j (-\chi_{1,i_p} \,y_1^{-1} )
\label{det00}
\ee
The first equality follows from the fact that $y_1(x)$ is the first eigenvalue of ${\bf g}$. One can easily build an $\epsilon$-deformation of this identity\footnote{This relation can be checked recursively by noticing that each term in the sum cancels the previous one leaving a
residue canceled by the next, and the process goes on and on and on. }
 \be
   \sum_{j=0}^r   \sum_{i_1<\ldots  <i_j}  \prod_{p=1}^j \left( - { \chi_{1,i_p} (x+p\epsilon-j \epsilon) \over y_1(x-p\epsilon+\epsilon)} \right)=0
   \label{det0}
 \ee
 which reduces to (\ref{det00}) when  $\epsilon=0$.
 On the other hand, the product in the right hand side of (\ref{det0}) can be written in terms of $\chi_j(x)$ using the
 identity
 \be
 \sum_{i_1<i_2 \ldots  < i_j} \prod_{p=1}^j \chi_{1,i_p}(x-j\epsilon+p\epsilon)  = \chi_j(x)\,\prod_{\ell=1}^{j} \tilde{Q}_{\ell}(x-(j-\ell+1)\epsilon)   \label{ident}
\ee
with
\be
\tilde{Q}_{\ell}(x)=\prod_{j=1}^{\ell-1}  Q_{j}(x)~~~~~~~~~~~~~~\ell=1,\ldots,r
\ee
 For $\epsilon=0$, $Q_a=1$, the relation (\ref{ident}) is nothing but the statement that the node $j$ in the $A$ Dynkin diagram is associated to the
 antisymmetric product of $j$ fundamentals.
 For $\epsilon$ and $Q_a$ generic the identity (\ref{ident}) can be checked using the explicit form
 of the characters. Replacing  $ \chi_j(x)$ by $\cP_j(x)$ in (\ref{ident}) one finds a non-trivial equation that is satisfied precisely
 at the solutions of the deformed Seiberg-Witten equations $ \chi_a(x)=\cP_a(x)$ , i.e.
 \be
 \framebox[1.15\width ][c]{$
   \sum_{j=0}^{r}  (-)^j  \, \cP_j(x)\,\prod_{\ell=1}^{j} {  \tilde{Q}_{\ell}(x-(j-\ell+1)\epsilon) \over y(x-(\ell-1) \epsilon)  }
      =0$}   \label{y1pol}
 \ee
 For $\epsilon=0$, we get the Seiberg-Witten curve
 \be
W_{A_{r-1} }(x,y)=\sum_{j=0}^{r}(-1)^{j} \,\cP_j(x)\prod_{\ell=1}^{j}(y^{-1}\,\tilde{Q}_{\ell})=0
\label{Wstand}
\ee
  In the rest of this section we show that the equation (\ref{y1pol}) can be thought as a quantum version of the Seiberg-Witten curve.
 Indeed  (\ref{y1pol}) can be found from (\ref{Wstand})   by  promoting the variables $x$ and $z=\log y$ to  non-commutative variables.   Explicitly, we replace $x$ and $z$ in (\ref{Wstand}) by the operators
   $\hat x$, $\hat z$   satisfying the non-commutativity relation
\be
[\hat{z},\hat{x}]=\epsilon
\ee
The quantum version of the Seiberg-Witten curve can then be written as
 \be
W_{A_{r-1} }(\hat{x}, \hat{y}=e^{\hat{z}})\,|\Psi\rangle=0
\ee
with the product over $\ell$ in (\ref{Wstand}) ordered in such a way that terms with greater $\ell$ go to the left.
Taking $\hat{z}=\epsilon\,\partial_{x}$, $|\Psi\rangle=\Psi(x)$ and using $e^{ \hat{z}} A(\hat{x})=A(\hat x+\epsilon)\,e^{ \hat{z}}$
we obtain
\be
\sum_{j=0}^{r}(-1)^{j}  \cP_j(x)\left(\prod_{\ell=1}^{j} \tilde{Q}_{\ell}(x-(j-\ell+1)\epsilon)\right)\,\Psi(x-j\,\epsilon)=0
\ee
After dividing by $\Psi(x-r\epsilon)$ and defining
\be
y(x)={\Psi(x)\over \Psi(x-\epsilon)}
\ee
we recover the deformed Seiberg-Witten curve (\ref{y1pol}). It would be nice to extend this analysis to the D and E series
and other representations of the A series.   The main difficulty in doing this comes from the fact that products of representations other than the fundamental leads to products of basic representations that require a more subtle ordering definition. We pospone a more systematic analysis of this issue to future investigations.
\\

 \subsection{A Thermodynamics Bethe ansatz  form}
 \label{satba}

 In this section we present an equivalent  form of the deformed Seiberg-Witten equations as a set of integral equations
 of TBA type. We first observe that
 the saddle point equations (\ref{eqsad}) imply that
 \be
1+   Q_a(x-\epsilon) {\prod_b y_b(x-m_{ba}-\epsilon  )^{c_{ba}} \,y_b(x +m_{ab} )^{c_{ab}}      \over  y_a(x-\epsilon) y_a(x) }
 ={\cP_a(x) \over y_a(x)} \Theta_a(x)   \label{tba0}
 \ee
 with $\Theta_a(x)$ a function with no zeros or poles around ${\cal I}_a$. Indeed the functions in the two sides
 of (\ref{tba0}) have zeros at $x_{u_a i}+\epsilon$ and poles at $x_{u_a i}$.   Taking the log of the two sides of this equation,
 multiplying by ${1\over z-x}$  and integrating around
 ${\cal I}_a$ one finds that  the $\Theta_a(x)$-dependent term cancels and one is left with the
 TBA integral equation
 \bea
&& \log y_a(z)= \log \cP_a(z)\label{tba}\\
&&~~~-\int_{\gamma_a}  {dx\over 2\pi i (x-z) } \log \left(    1+   Q_a(x-\epsilon) {\prod_b y_b(x-m_{ba}-\epsilon  )^{c_{ba}} \,y_b(x +m_{ab} )^{c_{ab}}      \over  y_a(x-\epsilon) y_a(x) }    \right) \nn
 \eea
 This equation can be easily solved order by order in $q_a$. In particular at one-instanton order one finds
 \be
 y_a(x)=  \cP_a(x)-    Q_a(x-\epsilon) {\prod_b \cP_b(x-m_{ba}-\epsilon  )^{c_{ba}} \,\cP_b(x +m_{ab} )^{c_{ab}}      \over  \cP_a(x-\epsilon) }
 \ee
 in agreement with the result (\ref{yya}) coming from the deformed Seiberg-Witten curve. We remark that function $\Theta_a(x)$ can be determined order by order in $Q_a$ by solving in $y_a$ these equations and requiring that $y_a$ has no poles
 outside of ${\cal I}_a$.

\vskip 1cm
\noindent {\large {\bf Acknowledgments}}
\vskip 0.2cm
The authors would like to thank R. Poghossian, V.Pestun and Y. Stanev for very useful discussions. This work is partially supported by the ERC Advanced Grant n.226455 ``Superfields" and by the Italian MIUR-PRIN contract 20075ATT78.

\vskip 1cm

\begin{appendix}

 \section{The affine case}
 \label{saffine}

 In this appendix we display the deformed Seiberg-Witten curves for the simplest quiver in the Affine series: the $A_1$ quiver.

\subsection{$A_1$ affine}

We consider here an Affine $A_1$-quiver gauge theory with $N_1=N_2=N$, no fundamental matter $n_1=n_2=0$
and two massless bifundamentals  $c_{12}=c_{21}=1$. Up to $k_1+k_2=3$ one finds
 \bea
 \chi_1(x)&=&y_1(x)+q_1\frac{y_2(x)\,y_2(x-\epsilon)}{y_1(x-\epsilon)}+\,q_1\,q_2\left(y_1(x)+\frac{y_1(x-2\epsilon)\,y_2(x)}{y_2(x-2\epsilon)}\right)+\ldots   \label{chis}   \\
&& q_1^2\,q_2\left(\frac{y_2(x)\,y_2(x-\epsilon)}{y_1(x-\epsilon)}+\frac{y_2(x)\,y_2(x-3\epsilon)}{y_1(x-3\epsilon)}\right)+\,q_1\,q_2^2\,\frac{y_1(x)\,y_1(x-\epsilon)\,y_1(x-2\epsilon)}{y_2(x-\epsilon)\,y_2(x-2\epsilon)}+\ldots\nn\\
 \chi_2(x)&=&y_2(x)+q_2\frac{y_1(x)\,y_1(x-\epsilon)}{y_2(x-\epsilon)}+\,q_2\,q_1\left(y_2(x)+\frac{y_2(x-2\epsilon)\,y_1(x)}{y_1(x-2\epsilon)}\right)\nn\\
&& q_2^2\,q_1\left(\frac{y_1(x)\,y_1(x-\epsilon)}{y_2(x-\epsilon)}+\frac{y_1(x)\,y_1(x-3\epsilon)}{y_2(x-3\epsilon)}\right)+\,q_2\,q_1^2\,\frac{y_2(x)\,y_2(x-\epsilon)\,y_2(x-2\epsilon)}{y_1(x-\epsilon)\,y_1(x-2\epsilon)}+\ldots\nn
 \eea
   The deformed Seiberg-Witten equations  are written as
    \be
  \chi_a(y,Q)=  \mathcal{P}_a(x)         \label{defaff}
\ee
with $\chi_a$ given by (\ref{chis}) and $g_a$ given by the leading term in a large $x$-expansion of the right hand side of
(\ref{chis}). One finds
 \be
g_a  = \frac{ \sum_{n\in{\mathbb Z}}q^{n^2-n}\,q_a^{n}}{(1-q) \prod_{n=1}^{\infty} (1-q^n) }  \label{gaff}
\ee
with $q=q_1 q_2$.

  \subsection*{The $\epsilon=0$ limit}

  In the limit $\epsilon=0$, the   right hand side of (\ref{chis}) sums up to
  \bea
  \chi_1(x)&=& { y_2(x)\over (1-q) \prod_{n=1}^{\infty} (1-q^n) }   \left(\frac{q_1}{q_2}\right)^{1\over 4}\,\vartheta_2(q_1\,{\ft {y_2(x)^2}{y_1(x)^2}}|q^2)\nn\\
   \chi_2(x)&=&  { y_2(x)\over (1-q) \prod_{n=1}^{\infty} (1-q^n) }   \,\vartheta_3(q_1\,{\ft {y_2(x)^2}{y_1(x)^2}}|q^2)
   \label{chitheta}
   \eea
   with
   \be
 \vartheta [^a_b ](y |q)=\sum_{n\in {\mathbb Z}} q^{{1\over 2}\left(n-{a\over 2} \right)^2}y^{n-{a\over 2} }e^{-\pi i b(n-{a\over 2})}
   \ee
   and $\vartheta_1= \vartheta [^1_1 ]$, $\vartheta_2= \vartheta [^1_0 ]$, $\vartheta_3= \vartheta [^0_0 ]$ and $\vartheta_4= \vartheta [^0_1 ]$. Using (\ref{defaff}) and (\ref{gaff}), the deformed Seiberg-Witten equations (\ref{chitheta}) can be  brought
    to the form
  \bea
  P_{e_1}(x)= y_2(x)\,\frac{\vartheta_2(q_1\,{\ft {y_2(x)^2}{y_1(x)^2}}|q^2)}{\vartheta_2(q_1|q^2)}\label{A11}\\
  P_{e_2}(x)= y_2(x)\,\frac{\vartheta_3(q_1\,{\ft {y_2(x)^2}{y_1(x)^2}}|q^2)}{\vartheta_3(q_1|q^2)} \label{A12}
  \eea
  with
  \be
  P_{e_a}=\prod_{u_a=1}^{N_a} (x-e_{u_a})
  \ee
    Interestingly, the ratio between these two equations depends only on the combination $y=y_1/y_2$. The function $y$ determines    the supergravity profile of the twisted field generated by the system of fractional branes at the $A_1$-singularity
    \cite{Billo:2012st}.
       Moreover in the limit   $q_2\to 0$ (where the gauge dynamics of the second node is turned off) (\ref{A11}) becomes simply $y_2(x)=P_{e_2}(x)$, while (\ref{A11}) leads to
  \be
y_1^2-y_1  P_{e_1}(x) \,(1+q_1)- q_1 P_{e_2}(x)^2   =0
  \ee
reproducing the Seiberg-Witten curve for $SU(N)$ gauge theories with $2N$ flavors.  The same result follows in the limit $q_1\to 0$.

\end{appendix}

\providecommand{\href}[2]{#2}\begingroup\raggedright\endgroup


\end{document}